# Photometric and Spectroscopic Analysis of the SX Phe Star BL Cam

Mohamed Abdel-Sabour, Mohamed I. Nouh, Ahmed Shokry, Gamal M. Hamed, Hamed A. Ismail, Ali Takey, Saad A. Ata and Ibrahim Zead

Astronomy Department, National Research Institute of Astronomy and Geophysics, 11421 Helwan, Cairo, Egypt

**mohamed.nouh@nriag.sci.eg**

**Abstract:** In the present paper, we report the photometric and spectroscopic observations obtained by the 1.88 m telescope at the Kottamia astronomical observatory of the pulsating star BL Cam. Fourier analysis of the light curves reveals that the fundamental mode has two harmonics. The O-C method is used to establish the period changes. So far, the analysis has been very successful in mapping the pulsation amplitude of the star across the instability strip. By using the formalism of Eddington and Plakidis (1929), we found significant results and strong indications of the evolutionary period change. A total of 55 new maximum light timings are reported. New values of (1/P) dP/dt are estimated using the O-C diagram based on all newly obtained times of maximum light combined with those taken from the literature, assuming the periods are decreasing and changing smoothly. To compute the effective temperature and surface gravity of the star, we performed model atmosphere analysis on its spectra. The physical parameters of the star are calculated and compared with the evolutionary models.

## 1. Introduction

SX Phoenicis (SX Phe) stars are typically found in the galaxy's outer regions, known as the galactic halo. Their luminosity changes over 1-2 hours. They are intriguing because they are very old but haven't turned into white dwarfs as expected. These stars exhibit short-period pulsation behaviour that varies on time scales ranging from 0.03 to 0.08 days (0.7 to 1.9 hours). SX Phe has spectral classifications in the A2-F5 range and magnitude differences of up to 0.7.

Giclas et al. (1970) discovered BL Cam (=GD 428 in Simbad, 2MASS J03471987+6322422, Gaia DR2 487276688415703040), which was thought to be a candidate for a white dwarf. It is a pulsating star with a period of 0.039 days and an amplitude of 0.33 mag, according to Berg & Duthie (1977). McNamara (1997) classified it as a Population II star with a metal abundance of [Fe/H] = 2.4. Previous researchers have looked into its multiperiodic character (Fauvaud et al. 2006; Rodrguez et al. 2007; Fu et al. 2008). Hintz et al. (1997) measured 32.679 c/d for the first overtone and 0.783 for the period ratio of the first overtone to the fundamental mode. Previous authors (Zhou et al. 1999; Kim & Sim 1999; Zhou et al. 2001; Wolf et al. 2002) had also discovered the initial overtone at 31.6 cd, resulting in a period ratio of 0.810. Fu et al., on the other hand, did not detect the first overtone (2008). In a multi-site photometric investigation of BL Cam, Rodrguez et al. (2007) discovered 21 distinct pulsation frequencies (excluding the basic mode) with amplitudes ranging from 1.6 to 7.4 mag.

According to Zong et al. (2019), the period content of BL Cam is dominated by 25.5790 (3) c/d and its two harmonics, as well as an independent frequency of 25.247 (2)c/d. An analysis of their times of maxima from the literature determined a periodic change which made BL Cam a binary system. Pena et al. (2021) has shown that the evolution of the ephemerides of the different authors was natural and correct given the shortness of the available data at their times. With a longer time, horizon, they demonstrated that the long-term variation is caused by a binary system.

In the present paper, we carried out photometric and spectroscopic observations for the star BL Cam. Model atmosphere analysis will be performed to determine the effective temperature and surface gravity of the star. Frequency analysis, O-C light curve, and period change of the star are investigated. The structure of the paper is as follows. The photometric analysis is presented in section 2, the spectral analysis is described in section 3, section 4 is devoted to the determination of the physical parameters and evolution state of the star, the conclusion reached is presented in section 5.

## 2. Photometric Analysis
## 2.1. Observation and Data Reduction

We present new photometric observations of BL Cam by using the 1.88m telescope of the Kottamia Astronomical Observatory (KAO), Egypt. Data Reduction and Photometry Bias subtraction and flat-field correction were applied to the raw CCD images without dark subtraction, which already was negligible. All observations were taken using EEV 42-40 CCD camera with a format of 2048*2048 pixels, cooled by liquid nitrogen to -120 C°. Figure 1 shows the field BL Cam taken with the KAO, the variable star, the comparison star, and the check star are marked as V, C1, and K respectively. Figure 2 shows the KAO observations obtained in Johnson BVR and the SDSS g, r, i, z filters for the three nights, 24, 26 Nov., and 25 Dec. 2021. All observations were analyzed using the MuniWin v.1.1.26 software (Hroch, 1998), implementing the differential magnitudes method of aperture photometry.

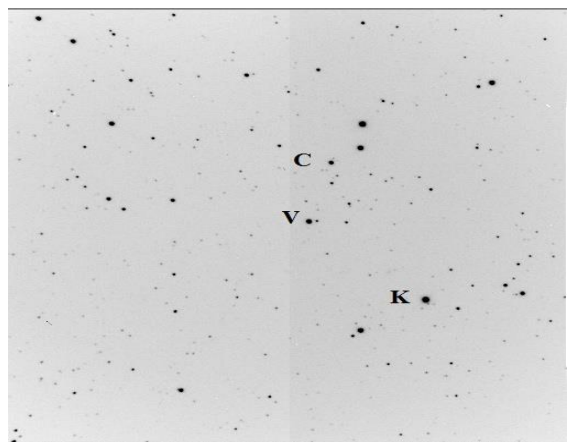

Figure 1. The Chart of the BL Cam, comparison (C), and check (K) stars.

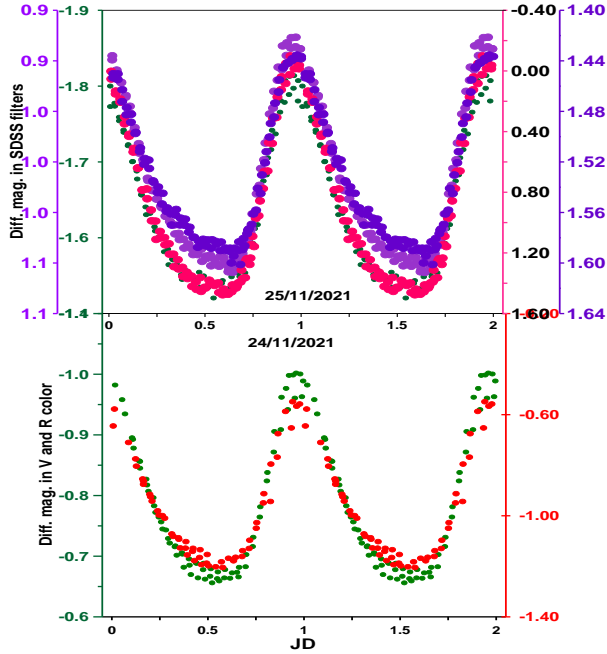

Figure 2: VR Color (lower panel) and g, r, i,z_color (upper panel) the differential magnitude of BL Cam.

**2.2. Frequency And Pulsation Analysis**

The frequency analysis of the BL Cam light curves was carried out with the help of two codes: PERIOD04 (Lenz and Breger 2005) and Peranso V3.0.3.4 (www.cbabelgium.com/peranso). Both codes searched for significant peaks in the amplitude spectra using Fourier transformations of the light curves. Following the first frequency computation, we create the "periodogram" by fitting a sinusoid to the Period04 period and then subtracting the sinusoid from the original magnitude (pre-whitening). Then we calculate the periodogram again, but at this moment the first frequency will not be presented, so the highest peak in the periodogram will be the next frequency. We repeated this procedure many times if necessary to search for other peaks until no more peaks could be seen in the periodogram.

Table 1 displays the results of all cases, including the frequency value (F), its corresponding amplitude (A), phase, and the signal-to-noise ratio (S/N). The V-band light curve analysis reveals two peaks in the periodogram at 0.03909787d (25.5768439 c/d) (Figure 3, upper panel) and 0.02922582 d (34.2163162c/d) (Figure 3, lower panel), with no identified peaks in the other bands. δ Scuti stars have a frequency range of 3-80 c/d, with frequencies less than 3 c/d possibly due to atmospheric effects or observational errors (Breger, 2000). The sum of the squared residuals ($\chi^2$) derived from a multi-parameter least squares fit of sinusoidal functions was used to calculate the error for each value. Figure 3 depicts the frequency spectra, Fourier fits on the observational points for all sets of observations and the spectral window of each star.

Table 1: The Frequency Analysis of BL Cam.

|    | Frequency (F) (c/d) | Amplitude (A) (mmag.) | Phase (deg) | S/N |
|----|---------------------|------------------------|-------------|-------|
| F1 | 25.5768439(2.31e-5) | 0.1298 (0.0178)        | 0.232(0.022)| 6.173 |
| F2 | 34.2163162(3.42e-5) | 0.0876(0.0178)         | 0.823(0.033)| 4.205 |
| F3 | 24.0362251(3.83e-5) | 0.0782(0.0178)         | 0.732(0.036)| 3.829 |

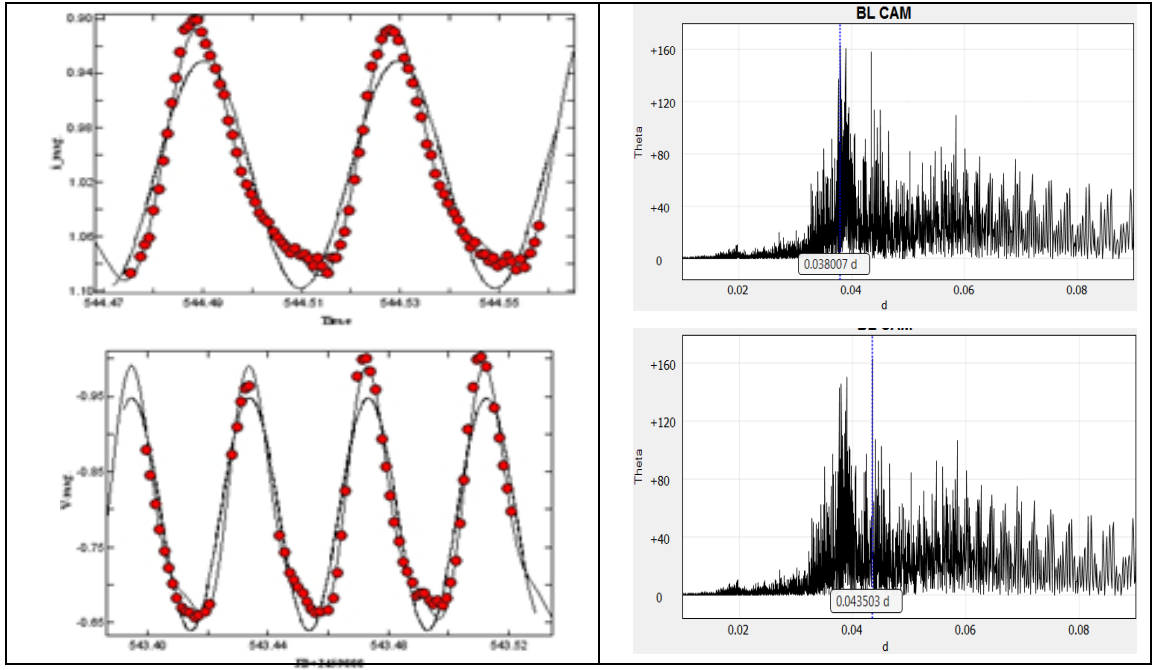

Figure 3: The amplitude spectra for BL Cam. The upper plots represent the i observations with the first frequency (25.5768439 c/d) while the lower plots are the V observations with the second frequency (34.2163162 c/d).

**2.3. O-C Light Curve and Period Change**

To construct the O-C light curve we used the Hertzsprung (1928) method to determine the time of brightness minima or maxima. We used all the data published in the literature to fill in the gaps in the O-C diagram, if the scatter is about the same as the scatter in the raw data (about 0.2 mag.) we use it. Figure 4 depicts all of the maximum-light times used to investigate the period change. To derive the O-C differences for BL Cam from a computed linear ephemeris, we used the method described by Abdel-Sabour et al. (2020). We take care to establish a reference time of maximum light from existing photoelectric observations, and the adopted pulsation period is based on recent observations of the star from KAO data. The least-square fit by the parabolic elements of the star under consideration is given by the following relationship:

$$HJD_{max} = M_0 + PE + QE^2.$$

Where $M_0$ is a new epoch, P is the new period, and Q is used to measure the period change values (dP/dt)

in seconds per year (dP/dt=(2Q/P)365.25x24x60x60).

To fit the O-C residuals, we used a second-order polynomial least-squares method. To justify its validity over all epochs of observations, the new ephemeris was tested with all available photometric observations from ASAS, KWS, and KAO for the star. For the light maximum, the linear ephemeris equation of Hintz et al. (1997) was used, which is:

$$HJD_{max} = 2443125.8026 + 0.03909783E$$

The 55 new times of maximum light obtained for BL Cam are presented in Table2. The Least square fit for O-C with root mean square $R^2 = 0.954$ is

$$HJD_{MAX} = 7.994(14) \times 10^{-2} - 5.0349(15) \times 10^{-8}E - 3.2757(69) \times 10^{-13}E^2,$$

The parabolic trend in O-C data reflects a regular period decrease or increase. After constructing the O-C diagram we found the increasing period rate given by equation dP/dt= 0.17028E-02 ±0.14378E-05 s/yr or $\frac{1}{P}(dP/dt) = 0.0436$ yr$^{-1}$, the standard deviation of the residuals of parabolic fit to the O-C values is $0^d.002$ with a correlation coefficient of 0.97.

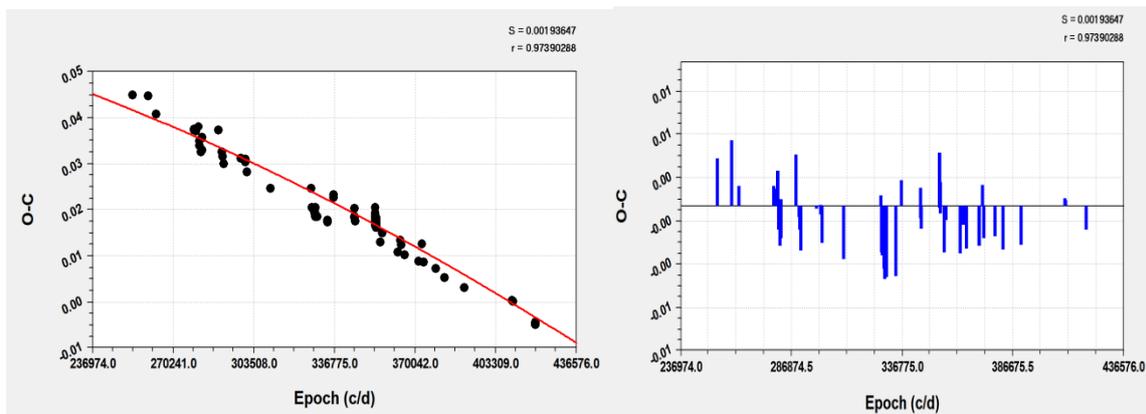

Figure 4: O-C data points fitted with a quadratic and the residuals of the quadratic fitting for BL Cam.

**Table 2:** New 37 times of maximum, new Epoch, O-C, number of observations, and the source of data.

| JD +2450000 | Epoch | O-C | No. of Obs. | Ref. | JD +2450000 | Epoch | O-C | No. of Obs. | Ref. |
|---|---|---|---|---|---|---|---|---|---|
| 3041.279 | 253607 | 0.04946 | 18 | 1 | 6186.419 | 334051 | 0.02209 | 60 | 1 |
| 3293.459 | 260057 | 0.04926 | 249 | 1 | 6186.458 | 334052 | 0.02170 | 56 | 1 |
| 3424.354 | 263405 | 0.04535 | 121 | 1 | 6280.259 | 336451 | 0.02717 | 35 | 1 |
| 4031.848 | 278943 | 0.04164 | 22 | 1 | 6280.298 | 336452 | 0.02776 | 40 | 1 |
| 4034.859 | 279020 | 0.04183 | 24 | 1 | 6623.336 | 345226 | 0.02268 | 262 | 1 |
| 4064.612 | 279781 | 0.04144 | 325 | 1 | 6623.375 | 345227 | 0.02287 | 220 | 1 |
| 4109.692 | 280934 | 0.04242 | 139 | 1 | 6623.416 | 345228 | 0.02483 | 255 | 1 |
| 4419.462 | 288857 | 0.04164 | 82 | 1 | 6630.372 | 345406 | 0.02189 | 201 | 1 |
| 4480.723 | 290424 | 0.03695 | 145 | 1 | 7034.402 | 355740 | 0.01740 | 28 | 2 |
| 4499.528 | 290905 | 0.03597 | 40 | 1 | 7074.440 | 356764 | 0.01935 | 44 | 1 |
| 4514.657 | 291292 | 0.03441 | 51 | 1 | 7314.652 | 362908 | 0.01525 | 58 | 2 |
| 4793.737 | 298430 | 0.03558 | 152 | 1 | 7362.901 | 364142 | 0.01779 | 48 | 1 |
| 4859.577 | 300114 | 0.03480 | 131 | 1 | 7370.719 | 364342 | 0.01681 | 48 | 1 |
| 4863.604 | 300217 | 0.03538 | 288 | 1 | 7437.457 | 366049 | 0.01466 | 37 | 2 |
| 4884.597 | 300754 | 0.03265 | 226 | 1 | 7651.398 | 371521 | 0.01329 | 37 | 2 |
| 5261.299 | 310389 | 0.02913 | 35 | 1 | 7715.639 | 373164 | 0.01701 | 133 | 1 |
| 5923.378 | 327323 | 0.02913 | 21 | 1 | 7745.935 | 373939 | 0.01310 | 47 | 2 |
| 5942.414 | 327810 | 0.02502 | 93 | 1 | 7942.947 | 378978 | 0.01173 | 41 | 2 |
| 5943.704 | 327843 | 0.02483 | 137 | 1 | 8077.636 | 382423 | 0.00977 | 39 | 2 |
| 5977.484 | 328707 | 0.02405 | 120 | 1 | 8394.755 | 390534 | 0.00762 | 35 | 2 |
| 5979.361 | 328755 | 0.02424 | 91 | 1 | 9168.884 | 410334 | 0.00469 | 60 | 1 |
| 5980.416 | 328782 | 0.02365 | 50 | 1 | 9177.642 | 410558 | 0.00450 | 93 | 1 |
| 5986.360 | 328934 | 0.02502 | 44 | 1 | 9543.434 | 419914 | -0.00059 | 20 | 3 |
| 5993.357 | 329113 | 0.02326 | 42 | 1 | 9543.473 | 419915 | -0.00039 | 17 | 3 |
| 5995.389 | 329165 | 0.02287 | 177 | 1 | 9543.512 | 419916 | -0.00020 | 27 | 3 |
| 5996.367 | 329190 | 0.02346 | 156 | 1 | 9543.512 | 419916 | -0.00020 | 7 | 3 |
| 6014.312 | 329649 | 0.02287 | 34 | 1 | 9544.529 | 419942 | 0.00000 | 90 | 3 |
| 6186.380 | 334050 | 0.02209 | 63 | 1 | | | | | |

1: All Sky Automated Survey (ASAS), 2-Kamogata/Kiso/Kyoto Wide-field Survey (KWS), 3: Kottamia Astr Observatory Astronomical.

## 3. Spectroscopic Analysis

We observed BL Cam using two grisms covering the spectral ranges 3360-5870 and 5300-9180, with spectral resolutions of 1025 in the blue part and 1133 in the red part. The spectra were taken with the Kottamia Faint Imaging spectropolarimeter (KFISP) mounted on the Kottamia Astronomical Observatory's 1.88 m telescope (KAO) (Azzam et al., 2021). The spectra were captured on a single night (25-11-2021). The data were reduced using the Astropy-affiliated package CCDPROC (Craig et al., 2017). We used van Dokkum and Pieter (2001) LACosmic routine to remove cosmic rays from the images processed by Astro-SCRAPPY (McCully et al., 2018). A special Python routine was used to extract the spectra and calibrate the wavelength. Using IRAF methods, the spectra were flux calibrated. The signal-to-noise ratio was calculated with specutils snr derived function (Earl et al., 2021). Table 3 shows the log of the spectroscopic observations. We plotted the BL Cam spectra in Figure 5. The upper panel represents the blue region, while the lower panel represents the red region.

Table 3: Observation Log of BL Cam Spectra

| DATE | TIME (UT) | JD | Phase | Exposure (s) | Standard Star | λ Range | R | Airmass | Average S/N |
|---|---|---|---|---|---|---|---|---|---|
| 25-11-2021 | 21:29:32.96 | 2459544.39554 | 0.376 | 900.0 | HR9087 | 3360-5870 | 1025 | 1.198 | 113.9 |
| 25-11-2021 | 21:59:40.35 | 2459544.41644 | 0.939 | 900.0 | HR9087 | 3360-5870 | 1025 | 1.206 | 113.3 |
| 25-11-2021 | 22:25:30.52 | 2459544.43438 | 0.406 | 900.0 | HR9087 | 3360-5870 | 1025 | 1.220 | 105.5 |
| 25-11-2021 | 23:01:03.12 | 2459544.45906 | 0.931 | 900.0 | HR9087 | 3360-5870 | 1025 | 1.252 | 108.0 |
| 25-11-2021 | 21:17:03.84 | 2459544.38687 | 0.598 | 600 | HR9087 | 5300-9180 | 1133 | 1.198 | 146.0 |
| 25-11-2021 | 21:48:45.29 | 2459544.40886 | 0.133 | 600 | HR9087 | 5300-9180 | 1133 | 1.202 | 166.2 |
| 25-11-2021 | 22:15:05.11 | 2459544.42714 | 0.592 | 600 | HR9087 | 5300-9180 | 1133 | 1.213 | 159.0 |
| 25-11-2021 | 22:44:36.40 | 2459544.44764 | 0.223 | 600 | HR9087 | 5300-9180 | 1133 | 1.235 | 189.6 |

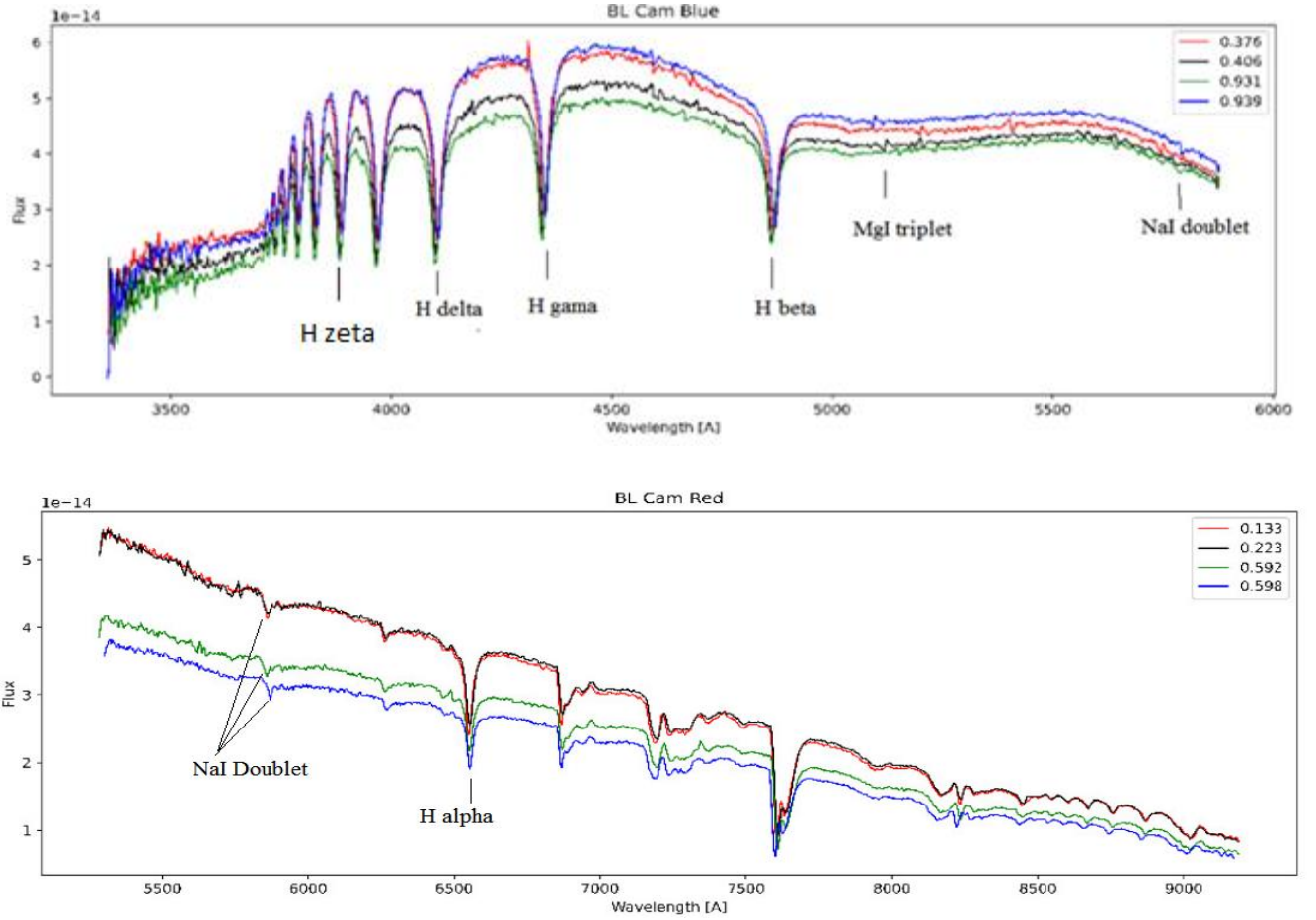

Figure 5: The observed spectra of BL Cam in the blue (upper panel) and red (lower panel) bands at different Phases.

We generated a small grid of synthetic spectra from LTE model atmospheres for the effective temperature range of $6000 \leq T_{eff} \leq 10000$ K for the spectral analysis. As input models for LTE calculations, we used ATLAS9 grids (Kurucz, 1995), assuming solar metallicity, a microturbulent velocity of 2 km/s, and a mixing length to scale height ratio of 1.25. The effective temperatures span the model grid's temperature range of 250 K. The surface gravities of the models are of $1 \leq \log g \leq 5$. The SPECTRUM code (Gray 1992, 1993) is used to synthesize LTE spectra (for the range $\lambda \lambda$ 1500 – 8000 Å). SPECTRUM takes the columns for mass depth points, temperatures, and total pressure and calculates them at each stage in the process using a system of seven nonlinear equilibrium equations.

We developed a FORTRAN code that compares the flux values at each point on the spectrum. It then tabulates the differences to produce a single number that characterizes how good the fit is. Because the spectral resolution of the observed spectra differs from that of the grid, the synthetic spectra are convolved to achieve the same resolution. For this purpose, a Gaussian profile with FWHM = 5 A° is used. We adjusted the wavelength scale to begin comparing the observed and grid spectra. We used the cross-correlation method to calculate the correlation coefficient, which gives us the line shifts. After that, we developed a code that minimizes the Euclidian distance between the observed spectra and the theoretical ones to compare the grid with the observed spectra. The equivalent widths of the spectral lines are calculated numerically using the Runge-Kutta technique.

Following the above procedure, we calculated the effective temperatures and the surface gravities of the star at different phases listed in Table 3. The mean effective temperature and surface gravity of the star is adopted as $T_{eff} = 7625 \pm 300$ K and $\log g = 4.30 \pm 0.37$ dex. Figure 6 shows the best fit of spectral lines at different phases in both red and blue parts of the spectrum. In most cases, we obtained a good fit for the line centers, while the large difference between the observed and the synthetic spectra occurs for the line wings.

Table 4 listed the equivalent widths of some spectral lines at different phases. The maximum variations of the equivalent widths with phases are 26%, 9%, 7%, 7%, and 25% for the lines $H_\alpha$, $H_\beta$, $H_\gamma$, $H_\varsigma$, NaI respectively. These variations raise the prospect that BL Cam is a binary star. We think that, high-resolution spectroscopic data is strongly needed to investigate the possibility of the binarity of the star.

Table 3: Effective temperatures and surface gravities at different phases of BL cam.

| Phase | $T_{eff}$ (K) | log g (dex) |
|---|---|---|
| 0.376 | 8000 | 4.0 |
| 0.406 | 8000 | 4.0 |
| 0.939 | 7750 | 4.0 |
| 0.931 | 7750 | 4.0 |
| 0.598 | 7250 | 4.5 |
| 0.592 | 7250 | 4.5 |
| 0.133 | 7500 | 4.5 |
| 0.233 | 7500 | 5.0 |
| Mean | 7625±300 | 4.30±0.37 |

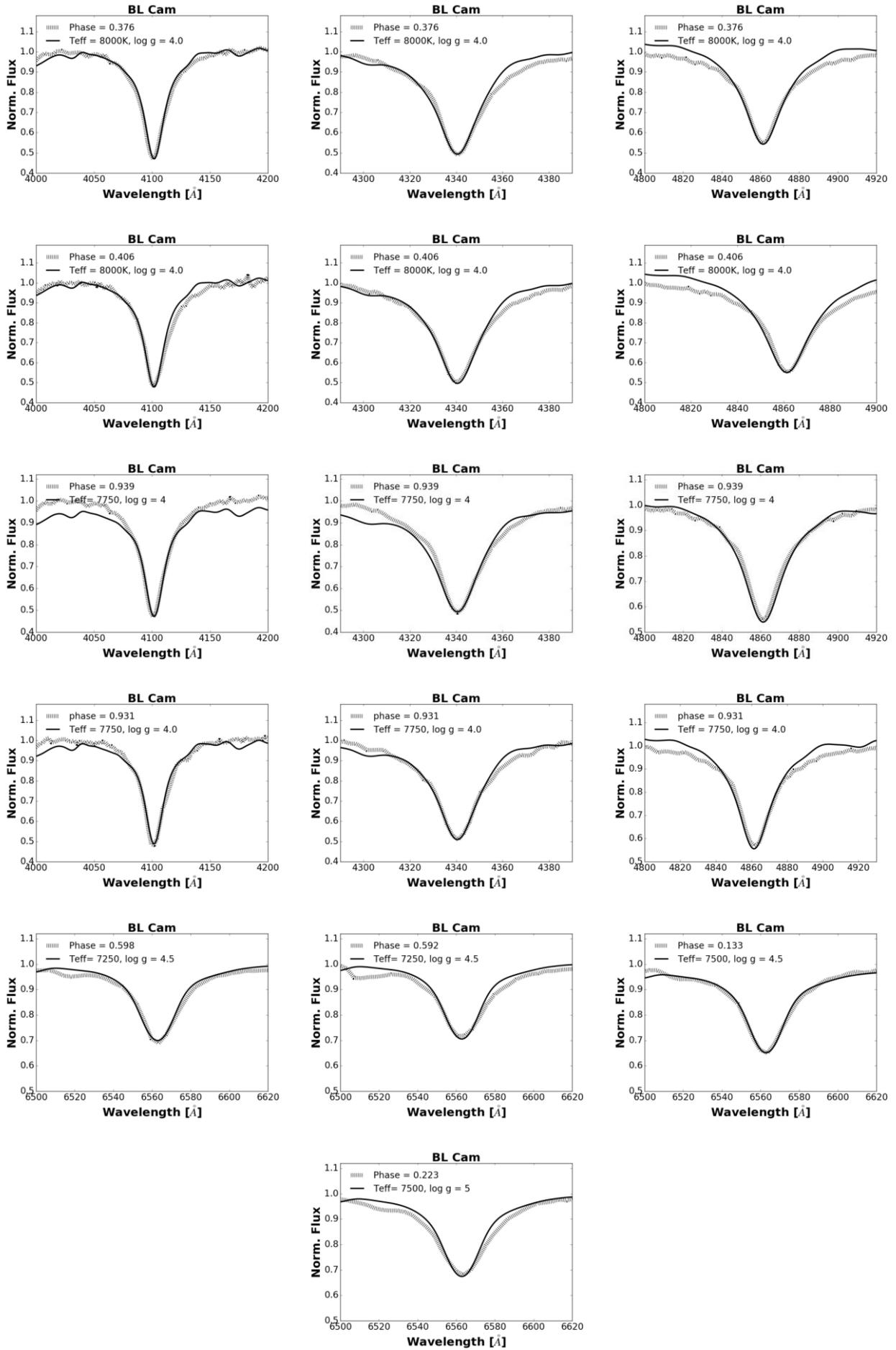

Figure 6: Comparison of observed Line profiles at different phases with that from synthetic spectra for the star BL Cam.

Table 4: The equivalent widths of some lines of the star BL Cam, a) for the blue part of the spectrum b) for the red part of the spectrum.

**(a)**

| Line/Phase | $W_\lambda$ (Å) | | | |
|---|---|---|---|---|
| | 0.376 | 0.406 | 0.931 | 0.939 |
| $H_\zeta$ (λ 3887 Å) | 11.07±0.08 | 11.08±0.17 | 10.31±0.37 | 10.26±0.05 |
| $H_\gamma$ (λ 4342 Å) | 13.60±0.73 | 14.59±0.7 | 14.56±0.48 | 14.14±0.46 |
| $H_\beta$ (λ 4862 Å) | 13.93±1.06 | 13.51±1.08 | 12.65±1.23 | 12.61±1.11 |

(b)

| Line/Phase | $W_\lambda$ (Å) | | | |
|---|---|---|---|---|
| | 0.133 | 0.223 | 0.592 | 0.598 |
| Na I (λ 5869 Å) | 1.08±0.11 | 1.04±0.08 | 0.82±0.12 | 1.09±0.18 |
| $H_\alpha$ (λ 6562 Å) | 13.04±1.73 | 12.01±0.83 | 9.56±1.87 | 8.76±0.47 |

## 4. Physical Parameters and the Evolution State

We computed the photometric physical parameters of BL Cam using the mean photometric colors and accurate parallax. Adopting the spectroscopic effective temperature as $T_{eff} = 7625 \pm 300$ K, and using the equation for the bolometric correction given by (Reed, 1998)

BC = -8.499 $[\log(T)-4]^4$ + 13.421$[\log(T_{eff})-4]^3$ - 8.131 $[\log(T_{eff})-4]^2$ - 3.901 $[\log(T_{eff})-4]$ - 0.438,

we evaluate the bolometric color as BC= -0.089, then the absolute magnitude of the star was calculated in the visible filter and the bolometric magnitude is calculated as $M_v$=3.426±0.061 and $M_{bol}$=3.337. The stellar radius is calculated from a polynomial fit to the temperature-radius relation of Gray (1992) as $R/R_\odot$= 1.494±0.032. The masses *M* can be calculated from the relation of Cox (1999) Log M= 0.46 -0.10 $M_{bol}$ as $M = 1.338 M_\odot$ Consequently the luminosity as $L/L_\odot$= 3.642. Using the frequencies in the periodogram (Table 1), the positional constant could be given by log Q=0.5 log g+0.1$M_{bol}$ +log $T_{eff}$ +log P-6.456,

Consequently, the Q value of f1 was 0.025(17), which was within the theoretical range for the fundamental mode given by Fitch (1981), where he found that pulsation constants lie in the range 0.0096 < Q < 0.067. The results are presented in Table 4. For the low-frequency pulsations ($v \leq 25d^{-1}$)), there is a good agreement with Bowman (2016).

In Figure 7, we plotted the mass-luminosity relation (M-L) and mass-radius relation (M-R) for both zero-age main-sequence stars (ZAMS) and terminal-age main-sequence stars (TAMS) with Z = 0.014 of Girardi et al. (2000). The positions of BL Cam on the two diagrams are close to the ZAMS track which indicates that it is an unevolved star.

Figure 8 (left panel) illustrated the position of BL Cam, as well as six SX Phe variables on the log Teff-log g diagrams for the masses, tracks 1.4 $M_\odot$-2 $M_\odot$. The parameters of BL Cam agree well with those predicted for SX Phe candidates listed in Table 6 adopted from Nemec and Mateo (1990). The tracks are plotted for the metallicity values Z=0.019 ([Fe/H] = -1.61) and Z=0.03 ([Fe/H] = -1.41). In the right panel of Figure 8, we plotted the isochrones appropriate for the effective temperature and luminosity of BL Cam, which indicated that the star has crossed the instability strip's red edge (RE). From this diagram, the age of BL Cam should be determined as 8.53±0.026 x$10^8$ years.

Table 5: Physical Parameters of selected Bl Cam.

| $T_{eff}$ (K) | M/$M_\odot$ | Log L/$L_\odot$ | R/$R_\odot$ | $M_{bol}$ | log g (dex) | Age(yr)x$10^8$ | Q (days) |
|---|---|---|---|---|---|---|---|
| 7625 | 1.68 | 0.957 | 1.69 | 2.335 | 4.3 | 8.53 | 0.025 |

Table 6: Effective temperatures and surface gravities of six SX Phe stars (Nemec and Mateo, 1990).

| Star name | $T_{eff}$ (K) | log g (dex) | Ref. |
|---|---|---|---|
| SX Phe | 7850 | 4.2 | 1 |
| KZ Hya | 7650 | 4 | 2 |
| CY Aqr | 7930 | 4.13 | 3 |
| BS Tuc | 7250 | 3.75 | 4 |
| DY Peg | 7800 | 4 | 3 |
| XX Cyg | 7530 | 3.66 | 5 |

Bessel (1969); 2) Przybylski and Bessell (1979); 3) McNamara and Feltz (1978); 4)Rodgers (1968); 5)Joner (1982).

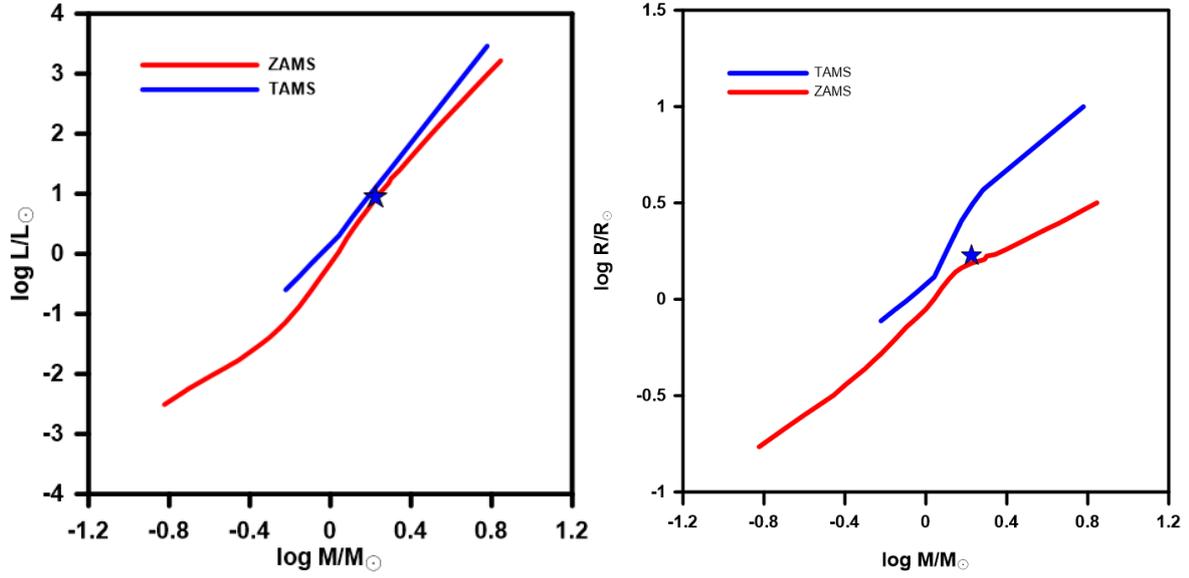

Figure 7: Position of BL Cam on the mass-luminosity (left) and temperature-radius (right) diagrams of Girardi et al. (2000) evolution models.

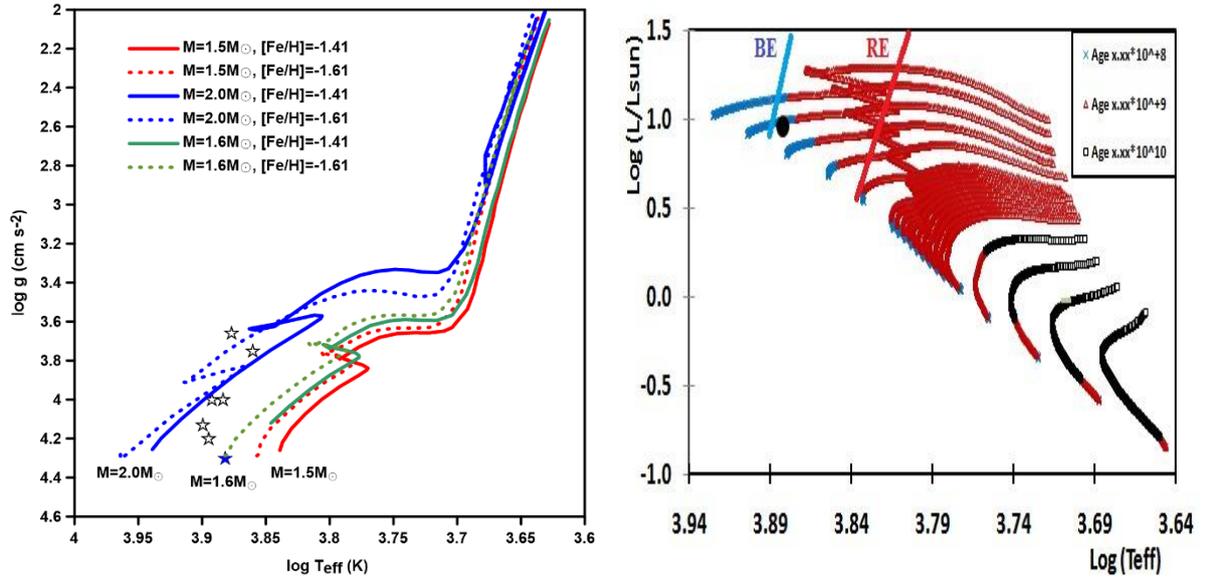

Figure 8: Position of BL Cam on the temperature-gravity (left), temperature-luminosity (right) of Girardi et al. (2000) evolution models for the metallicity Z=0.03 (solid lines) and Z=0.019 (dashed lines).

## 5. Conclusion

We conducted a thorough analysis of the star BL Cam using photometric and spectroscopic observations obtained at the Kottamia observatory. According to the Fourier analysis of the light curves, the fundamental mode has two harmonics at 25.5768439 c/d and 34.2163162 c/d. We combined the new times of maximum light with those provided by previous literature to perform an O-C analysis for the period change for BL Cam, yielding 55 times of maximum light. The variation rate of the fundamental period derived from the long time scale of observations shows a negative period change.

We used LTE model atmospheres to simulate the observed spectra and to calculate equivalent widths of some spectral lines. By comparing the spectra to the appropriate synthetic spectra, the effective temperatures and surface gravities at different phases are calculated. We adopted the effective temperature and surface gravity of BL Cam as $T_{eff} = 7625 \pm 300$ K and $\log g = 4.30 \pm 0.37$ dex, which is in good agreement with earlier studies. Our analysis of the spectra of BL Cam indicated that the variations of the equivalent widths of some spectral lines at different phases are 26%, 9%, 7%, 7%, and 25% for the lines $H_\alpha$, $H_\beta$, $H_\gamma$, $H_\zeta$, NaI respectively. These variations raise the possibility that BL Cam is a binary star; but to follow this suggestion, we need spectroscopic observation with reasonable resolution.

We located the physical parameters of the star on the evolutionary models to investigate its evolution state. The calculated mass (1.68 $M_\odot$) is in a good agreement with mass tracks around 1.6 $M_\odot$ and higher than the possible masses (1.0$M_\odot$- 1.4$M_\odot$) of the SX Phe stars predicted by Nemec and Mateo (1990).